\documentclass[aps,twocolumn,superscriptaddress,nobibnotes]{revtex4-1}

\usepackage{mathtools}
\usepackage{array}
\usepackage{amsmath}
\usepackage{cancel}
\usepackage{float}

\begin{document}

\renewcommand{\vec}[1]{\mathbf{#1}}
\let\oldhat\hat
\renewcommand{\hat}[1]{\oldhat{\mathbf{#1}}}


\title{Compact interaction potential for van der Waals nanorods}


\author{Jack A. Logan}
\affiliation{Department of Physics and Astronomy, Stony Brook University, Stony Brook, NY 11794, USA}
\affiliation{Center for Functional Nanomaterials, Brookhaven National Laboratory, Upton, New York 11973, USA}

\author{Alexei V. Tkachenko}
\affiliation{Center for Functional Nanomaterials, Brookhaven National Laboratory, Upton, New York 11973, USA}


\date{\today}

\begin{abstract}
We studied the van der Waals interactions of two finite, solid, cylindrical rods at arbitrary angle and position  with respect to each other. An analytic interpolative formula for the  interaction potential energy is constructed, based on various asymptotic cases. The potential can be readily used for numerical and analytic  description of multi-wall carbon nanotubes, metallic nanorods, rod-shaped colloids, or any other similar objects with significant van der Waals interactions. 
\end{abstract}


\maketitle

\section{Introduction \label{sec:Introduction}}

Often thought of as short-ranged and weak, van der Waals (vdW) interactions  in fact  play a profound role in our everyday life, and in a variety of research fields, from materials science to chemistry and biology \cite{israelachvili2011intermolecular,parsegian2005van}. For instance, vdW forces are responsible for stacking of layers in graphite and other heterostructures \cite{geim2013van, lopez2014light},  for protein stability \cite{nick2014forces}, for various types of self-assembly \cite{shimoda2002self} adhesion  and capillary phenomena, and even for a Gecko's remarkable ability to stick to any surface \cite{autumn2002evidence,autumn2000adhesive}. These interactions become especially important on micro- and nanoscales. Second in strength only to electrostatics, it often drives aggregation of nanoparticles and colloids \cite{israelachvili2011intermolecular,parsegian2005van}.

The theory behind  vdW interactions was first developed by London \cite{london1937general}, and  interpreted as an effect of correlated quantum fluctuations of  dipole moments \cite{israelachvili2011intermolecular,parsegian2005van,hamaker1937london,lifshitz1956theory}. When retardation and multi-body effects are neglected, it is described as a simple $1/r^6$ potential. This potential can  be integrated for a variety of shapes  of the interacting objects, such as lines, flat surfaces, spheres, etc \cite{israelachvili2011intermolecular,parsegian2005van,hamaker1937london}. In this paper we focus on  a particular case of rod-like particles with vdW interactions. It is motivated by an important role that rod-shaped objects play in modern nanoscience. They include carbon nanotubes, metallic nanorods, microtubules and others \cite{sainsbury2005experimental,stolarczyk2007evaluation,kamal2009van,rance2010van, liu2017assembling,huang2009gold}.

The problem has been partially addressed in the past, primarily  in the context of carbon nanotubes \cite{ruoff1993radial,tersoff1994structural,girifalco2000carbon,zhbanov2010van,pogorelov2011universal}. The key challenge is that, although exact numerical  integration of vdW potential for a pair of rods is certainly possible, the result is not expressed as a compact closed form. This makes it very problematic to use such calculation as  a part of  e.g. multi-particle simulation. In this paper, we present a compact closed form description of vdW rods which is not exact, but a high accuracy interpolation in a multi-parameter space. Specifically, we consider two uniform rods of the same length $L$, diameter $a$, and give an approximate formula for vdW potential as function of their relative positions and orientation. Together with $L$ and $a$, there are $6$ independent parameters in this problem.





\section{Results \label{sec:Results and Discussion}}

\subsection{Model \label{sec:Model}}



\begin{center}
\begin{figure}[hbtp]
\includegraphics[width=1\linewidth]{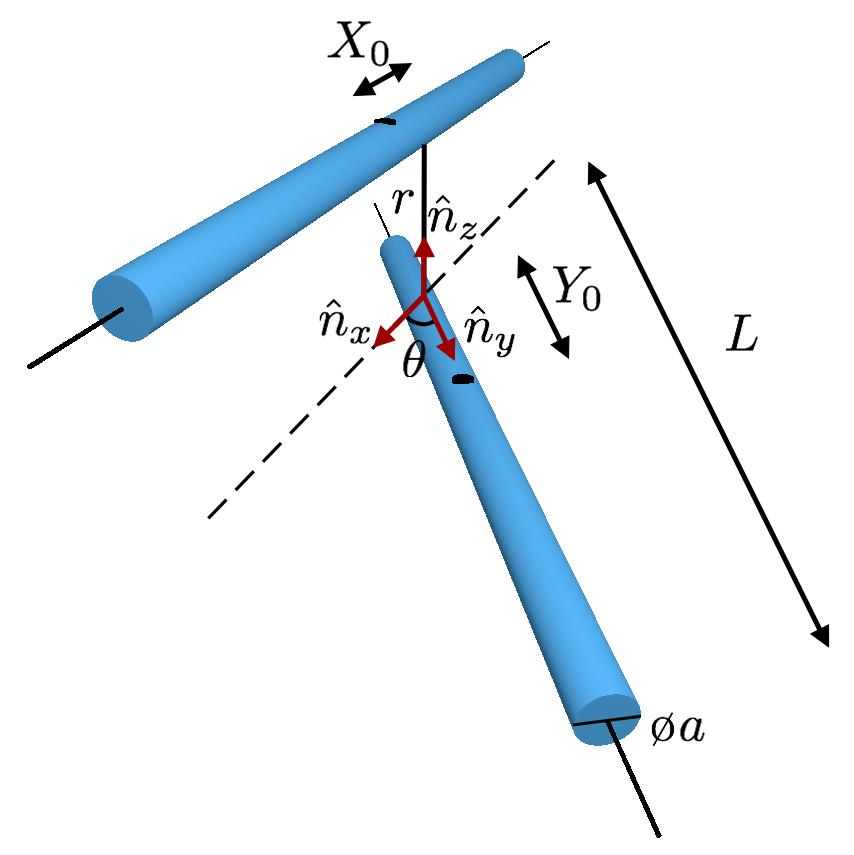} 
\caption{Basic setup:  two identical rods of diameter $a$ and  length $L$.   $X_0$ and $Y_0$ are longitudinal displacements of the rods' centers with respect to their axes' nearest points.}
\label{fig:Model}
\end{figure}
\end{center}

The goal of this paper is to propose a simplified, but accurate, model potential for the vdW attractive energy of two finite rods.   Our system  is shown  in Fig.~\ref{fig:Model}. Unit vectors $\hat{n}_x$ are $\hat{n}_y$ are directed along the two rods,   and $\hat{n}_z$ is perpendicular to both of them. In addition to the  ``director" vectors of the two rods, $\hat{n}_x$ are $\hat{n}_y$, we assume that the positions of their centers, $\vec{X}_c$ and $\vec{Y}_c$ are given as well:

\begin{align*}
\vec{X}_c &=\vec{X}_1 +X_0 \, \hat{n}_x \\ \vec{Y}_c &= \vec{Y}_1+Y_0 \, \hat{n}_y, 
\end{align*}

Here $\vec{X}_1$ and $\vec{Y}_1$ are the two closest points that belong to the axes of the two rods, respectively.  $X_0$ and $Y_0$ are their longitudinal displacements (see Fig \ref{fig:Lines}),  

\begin{equation*}
X_0 = \frac{\left[ \left( \vec{X}_c - \vec{Y}_c \right) \cdot \hat{n}_y \right]\left( \hat{n}_x \cdot \hat{n}_y \right) - \left( \vec{X}_c - \vec{Y}_c \right) \cdot \hat{n}_x}{\left( \hat{n}_x \cdot \hat{n}_y \right)^2 -1}
\end{equation*}

\begin{equation*}
Y_0 = \frac{\left[ \left( \vec{X}_c - \vec{Y}_c \right) \cdot \hat{n}_x \right]\left( \hat{n}_x \cdot \hat{n}_y \right) - \left( \vec{X}_c - \vec{Y}_c \right) \cdot \hat{n}_y}{1 - \left( \hat{n}_x \cdot \hat{n}_y \right)^2}.
\end{equation*}

The axis-to-axis distance between the rods is 

\begin{equation*}
r=\left|\left(\vec{X}_c - \vec{Y}_c\right)\cdot \hat{n}_z\right|,
\end{equation*}

where

\begin{equation*}
\hat{n}_z = \mathrm{sgn} \left( \left(\vec{X}_c-\vec{Y}_c \right)\cdot \left( \hat{n}_x \times \hat{n}_y \right) \right) \frac{\hat{n}_x \times \hat{n}_y}{\left| \hat{n}_x \times \hat{n}_y \right|}.
\end{equation*}

With this definition of $\hat{n}_z$, the origin of the $XYZ$ system is at point  $\vec{Y}_1$, and vector $\hat{n}_z$  points towards the ``X" rod.  The angle between two rods is defined as 

\begin{equation*}
\sin \theta = \left| \hat{n}_x \times \hat{n}_y \right|.
\end{equation*}

The attractive vdW energy for two objects (in our case, rods) can be calculated as


\begin{equation}\label{eq:Vol Interaction}
U = \frac{-A}{\pi^2} \int \int \frac{d^3 \vec{r}_1 \, d^3 \vec{r}_2}{|\vec{r}_1-\vec{r}_2|^6}.
\end{equation}

Here $A$ is a material-dependent Hamaker constant,  and integration over $\vec{r}_1$ and $\vec{r}_2$ is carried out within each of the objects, respectively. 



The plan of this paper is as follows. In the next section, we explore the limit of long rods, without account for any effects of terminals. 
We first construct the potential in the near- and far-field limits, both for parallel, and non-parallel. The near-field regime is defined as $(r-a)\ll a$, the  the far-field as $r \gg a$. The  near-parallel orientation corresponds to  $\sin \theta \ll \frac{a}{L}$, and the non-parallel is the opposite limit: $\sin \theta \gg \frac{a}{L}$. From the four limiting cases, we then construct a single interpolative formula.  

In the following  section,  we discuss the effects of rods' ends    in the far-field approximation, both  for parallel and non-parallel rods. Finally,  by  combining  results from both  sections, a unified formula will be obtained.






\subsection{Infinite Length with Finite Diameter \label{sec:Infinite_Length_with_Finite_Diameter}}

\subsubsection{Non-Parallel Rods \label{sec:Non-Parallel Rods finite diameter}}



\begin{figure}[hbtp]
\begin{center}
\includegraphics[width=1\linewidth]{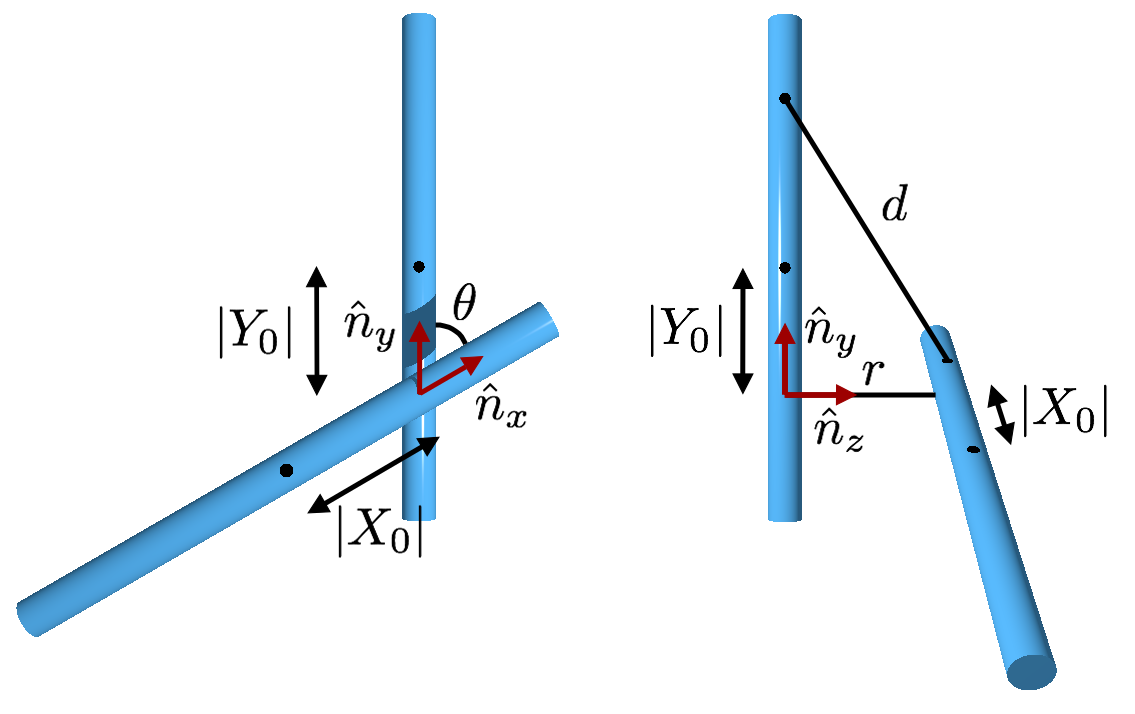} 
\end{center}
\caption{Two views of a pair of non-parallel rods.  }
\label{fig:Lines}
\end{figure}

We start by considering infinitesimally thin rods, which corresponds to the far-field regime, $a\ll r$. Let $(X,0,r)$ and $(0,Y,0)$ be two points that belong to the two different rods (in  our non-orthogonal  coordinates  $(\hat{n}_x,\hat{n}_y,\hat{n}_z)$). As seen from  Fig.~\ref{fig:Lines}, the distance, $d$, between these points, can be found as \newline $d^2=r^2 + (X\sin\theta)^2 + (Y-X\cos\theta)^2$.

With the substitution $x=\frac{X\sin\theta}{r}$ and $y=\frac{Y-X\cos\theta}{r}$, the potential becomes

\begin{equation}\label{eq:AttrInt}
U=\frac{-A \, \sigma_1 \sigma_2}{\pi^2 \, r^4 \left| \sin \theta \right|} \int_{x_1}^{x_2} \int_{y_1}^{y_2} \frac{dy \, dx}{(1+x^2+y^2)^3}, 
\end{equation}
where $\sigma_1=\sigma_2=\frac{\pi a^2}{4}$ is the cross-sectional area of each rod. In the limit of infinitely long rods,  we obtain:

\begin{equation}\label{eq:Uinf}
U_{\times}^{\infty} = \frac{-A \pi a^4}{32 \left| \sin\theta \right| r^4}
\end{equation}

Here the superscript reminds us that this is a far-field result, and the subscript shows that the rods are non-parallel. We can use this thin-rod result to calculate the vdW potential of two finite-thickness rods: 
\begin{equation}
U_{\mathrm{\times}} = \int \int \frac{-A \, d\sigma_1 d\sigma_2}{2 \pi \left| \sin \theta \right| (\vec{z}_1 - \vec{z}_2)^4} 
\end{equation}
Here integration is carried out over the cross-section areas of each rod,  $z_1$ and $z_2$ are $z$  coordinates of the  respective area elements.   The result of this integration can be expressed as a simple formula in the near-field regime ($(r-a)\ll a$), which complements the above far-field result:

\begin{equation}\label{eq:Urods,limits}
U_{\mathrm{\times}} \approx \begin{cases}{} 
\frac{-A \pi a^4}{32 \,  \,\left|\sin\theta\right| \, r^4} & \frac{r}{a} \gg 1 \\ \\ 
\frac{-A a}{12\,  \,\left|\sin\theta\right| \, (r-a)} & \frac{r-a}{a} \ll 1
\end{cases} 
\end{equation}



\vspace{5mm}

In order to see how the energy transitions from the near-field to the far-field, $U_{\mathrm{\times}}$ is  evaluated exactly using numerical integration. A fit of the exact solution using Eq.~\ref{eq:Urods,limits} was interpolated with

\begin{equation*}
U = \frac{C}{(r-a)(r+B)^3}.
\end{equation*}

Using the asymptotes, we can find $B$ and $C$ such that the equation has the correct $r$-dependence in the two limiting cases:



\begin{equation}\label{eq:Urods fit w/o epsilon}
U = \frac{-V_0}{ \left| \sin \theta \right| \left(r-a \right) \left(r + \left(\frac{1}{2}\left( 3 \pi \right)^{\frac{1}{3}}-1 \right)a \right)^3},
\end{equation}
Here  $V_0 = \frac{A \pi a^4}{32}$.

\begin{center}
\begin{figure}[h]
\includegraphics[width=1\linewidth]{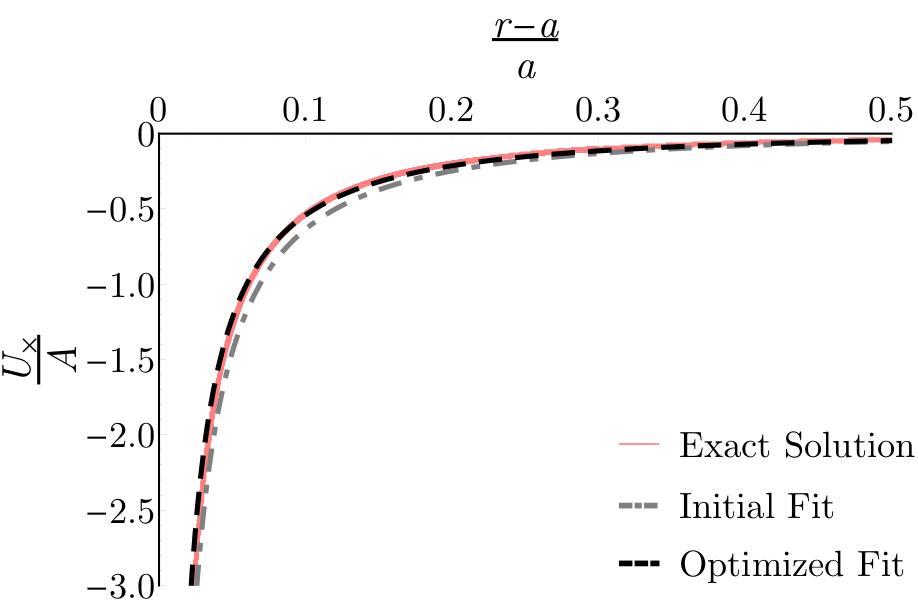} 
\caption{Comparison of the exact numerical solution for  skew rods with finite diameter (solid red), and the approximations given by Eq.~\ref{eq:Urods} (dashed black) and Eq.~\ref{eq:Urods fit w/o epsilon} (dot-dashed gray).}
\label{fig:Urods fit}
\end{figure}
\end{center}

This is a good fit in the near- and far- fields, but it slightly deviates from the exact in the intermediate range. We found that the prefactor of $a$ in Eq.~\ref{eq:Urods fit w/o epsilon}, $\left( \frac{1}{2}\left( 3 \pi \right)^{\frac{1}{3}}-1 \right) \approx 0.06$, can be used as a fine-tuning parameter $\epsilon$, and a near-perfect fit is achieved for $\epsilon=0.12$, as shown in Fig.~\ref{fig:Urods fit}, 


\begin{equation}\label{eq:Urods}
U_{\mathrm{\times}} = \frac{-V_0}{\left| \sin \theta \right| (r-a)\left(r+\epsilon \, a\,\right)^3}
\end{equation}



\vspace{5mm}


\subsubsection{Parallel Rods \label{Parallel Rods with finite diameter}}

Next, we would like to see what happens when finite-diameter rods become parallel. The interaction for infinite length rods in the far-field can be found using Eq.~\ref{eq:Vol Interaction}, and the near-field can be calculated with the help of Derjaguin's approximation \cite{israelachvili2011intermolecular}. Together, the near- and far-field results for parallel rods are given by 



\begin{equation}\label{eq:Urods parallel, limits}
U_{\mathrm{\parallel}} \approx \begin{cases}{} 
\frac{-3\pi A a^4 L}{128 r^5} & \frac{r}{a} \gg 1 \\ \\ 
\frac{-A L \sqrt{a}}{24\sqrt{2} (r-a)^{3/2}} & \frac{r-a}{a} \ll 1.
\end{cases} 
\end{equation}

We start with Eq.~\ref{eq:Urods}, which we know works well in both the near- and far-fields for non-zero angles, but replace $\left|\sin \theta \right|$ with a correction term to match the parallel-rods results in both limits,  Eq.~\ref{eq:Urods parallel, limits}:

\begin{equation}\label{eq:Urods approx, parallel}
U_{\mathrm{\parallel}} \approx \frac{-V_0}{\beta \frac{\sqrt{r(r-a)}}{L} (r-a)(r+\epsilon a)^3}
\end{equation}

The exact result for parallel  rods of finite diameter  can be obtained by numerical integration of the thin-rod (far-field) potential: 

\begin{equation}\label{eq:Exact, parallel rods}
U_{\mathrm{\parallel}} = \int \int \frac{-3 A L \, d\sigma_1 d\sigma_2}{8 \pi \, r_{12}^5}.
\end{equation}

Here $r_{12}$ is the distance between respective area elements.

As shown in Fig.~\ref{fig:Urods fit, parallel}, the result of this integration is in excellent agreement with our interpolation formula, Eq.~\ref{eq:Urods approx, parallel}. The best fit is achieved for    $\beta=2.35$. 

\begin{center}
\begin{figure}[h]
\includegraphics[width=1\linewidth]{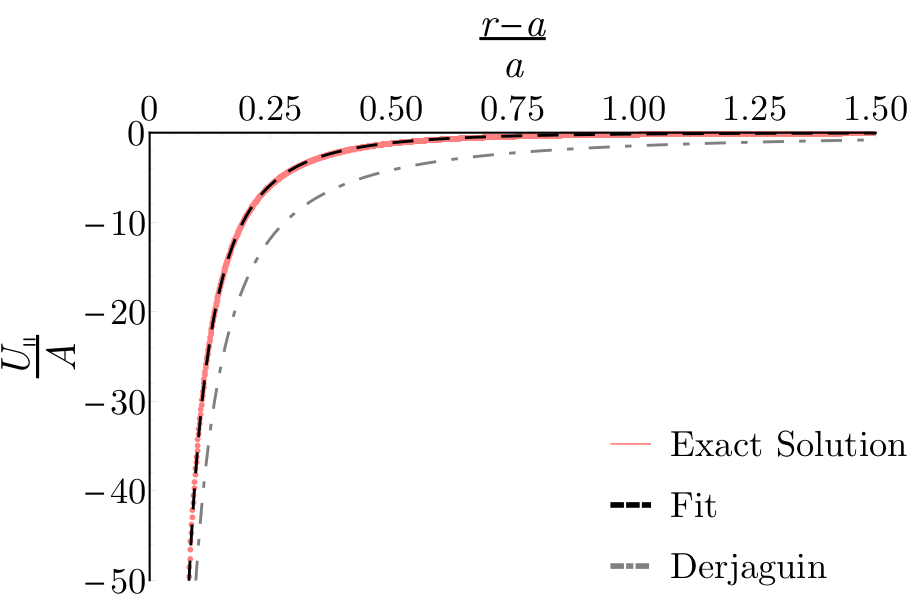} 
\caption{Comparison of the exact numerical solution of parallel rods with finite diameter (solid red), and the approximations given by Eq.~\ref{eq:Urods approx, parallel} with $\beta=2.35$ (dashed black) and the Derjaguin approximation (dot-dashed gray).}
\label{fig:Urods fit, parallel}
\end{figure}
\end{center}

By combining $U_\times$ and $U_\parallel$, the long-rod  result for arbitrary  angle and diameter can be obtained:

\begin{equation}\label{eq:Urods, final}
U_{\mathrm{rods}} \approx \frac{-V_0}{(\left|\sin \theta \right| + 2.35 \frac{\sqrt{r(r-a)}}{L}) (r-a)(r+\epsilon a)^3}.
\end{equation}

\subsection{Finite Length in the Far-Field \label{sec:Finite Rods}}

\subsubsection{Orthogonal Rods \label{sec:Far-Field Orthogonal Rods}}

In the previous section we found an accurate solution for long rods with finite diameters, but without account for any effects related to rod terminals. Below we explore how the proximity of the rods' ends alter the above result. 


First, we use Eq.~\ref{eq:AttrInt}, with $\theta = \frac{\pi}{2}$ and limits $x_1=y_1=-\infty$, $y_2=\infty$, and $x_2=\frac{X}{r}$, to find the interaction, $U_{\mathrm{s}}$, between an infinite rod and a semi-infinite rod perpendicular to each other. 

\begin{equation}\label{eq:U inf w/ semi-inf}
U_{\mathrm{s}} = U_{\times}^{\infty} \left[\frac{3 \frac{X}{r}+2(\frac{X}{r})^{3}}{4(1+(\frac{X}{r})^{2})^{3/2}}+\frac{1}{2}\right]
\end{equation}

\begin{center}
\begin{figure}[H]
\includegraphics[width=1\linewidth]{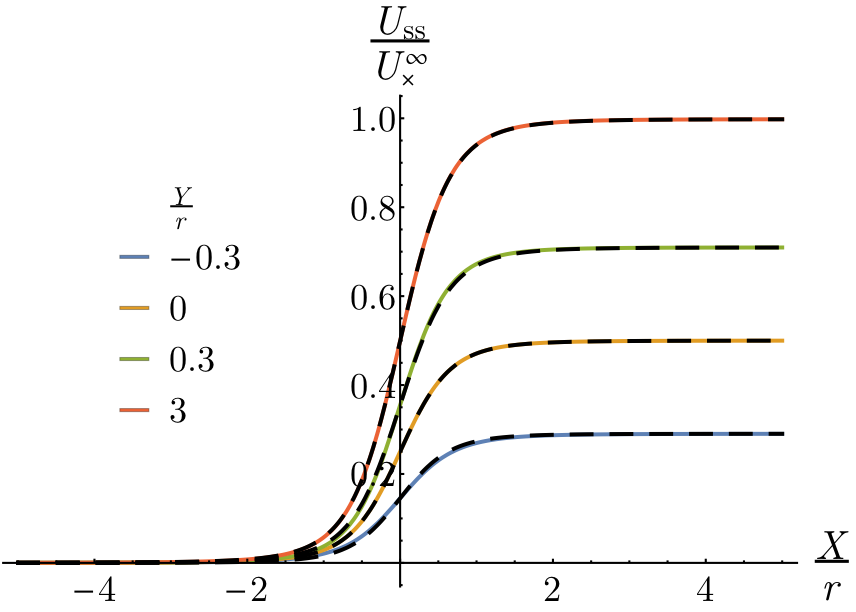} 
\caption{Comparison of $\frac{U_{\mathrm{ss}}}{U_{\times}^\infty}$ with the approximation from Eq.~\ref{eq:Uss,app} (dashed black) formed from $\frac{U_{\mathrm{s}}}{U_\times^\infty}$, with $\frac{L}{r}=25$.}
\label{fig:Uss approx and exact}
\end{figure}
\end{center}

We next evaluate the exact solution for two perpendicular semi-infinite rods, $U_{\mathrm{ss}}$, with limits $x_1=y_1=-\infty$, $x_2=\frac{X}{r}$, and $y_2=\frac{Y}{r}$, using Eq.~\ref{eq:AttrInt} and observe that a factorized formula based on Eq.~\ref{eq:U inf w/ semi-inf} is a good approximation, as shown in Fig.~\ref{fig:Uss approx and exact}.


\begin{equation}\label{eq:Uss,app}
\frac{U_{\mathrm{ss}}}{U_{\times}^{\infty}} \approx \left[G \left(\frac{X}{r} \right)+\frac{1}{2}\right] \left[G \left(\frac{Y}{r} \right)+\frac{1}{2}\right], 
\end{equation}

with \begin{equation*}
G(x) = \frac{3 x +2 x^{3}}{4\left(1+x^{2}\right)^{3/2}}.
\end{equation*}

Equation~\ref{eq:Uss,app} is a good fit for two semi-infinite rods, and we can modify it to make the rods finite. Using Eq.~\ref{eq:Uss,app} we define a function $\Gamma$ as


\begin{equation}\label{eq:gamma}
\Gamma(x_\pm, y_\pm) = \left[G\left(x_+ \right)-G\left(x_-\right) \right] \, \left[G\left(y_+ \right)-G\left(y_-\right) \right].
\end{equation}
Here  $x_\pm=\frac{1}{r} (X_0 \pm \frac{L}{2})$, $y\pm=\frac{1}{r} (Y_0 \pm \frac{L}{2})$. 
This factor gives a perfect description of the finite-size correction to our long-rod result  for perpendicular rods:
\begin{equation*}
U = U_{\times}^{\infty}\Gamma\left( \frac{X_0 \pm \frac{L}{2}}{r}, \frac{Y_0 \pm \frac{L}{2}}{r}\right).
\end{equation*}
However, as the angle $\theta$ between rods  changes, this correction becomes increasingly inadequate, especially in the limit of parallel rods. 

As a remedy, we introduce an alternative, quasi-linear corrective factor:
\begin{equation}
\gamma(x_\pm, y_\pm) = \mathrm{min}\left\{\left[g\left(x_+ \right)-g\left(x_-\right) \right], \left[g\left(y_+ \right)-g\left(y_-\right) \right]\right\},
\end{equation}
with $g(x) = \frac{1}{2} \, \mathrm{sgn}\left(x\right) \, \mathrm{min}\left\{ 1, \frac{3}{2}\left|x\right| \right\}$.
The advantage of this function is two-fold: first, it is much simpler to evaluate $\gamma$ and its derivatives (which is needed for finding forces); second, it can be easily modified to describe the parallel rod limit. 

\begin{center}
\begin{figure}[H]
\includegraphics[width=1\linewidth]{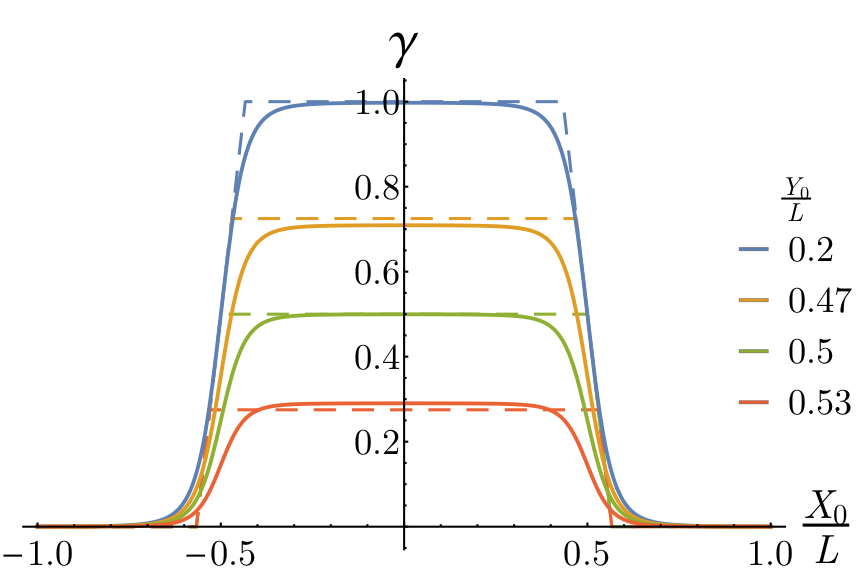} 
\caption{Comparison of two versions of finite-size factors, $\gamma(\frac{X_0\pm\frac{L}{2}}{r}, \frac{Y_0\pm\frac{L}{2}}{r})$ (dashed), and  $\Gamma(\frac{X_0\pm\frac{L}{2}}{r}, \frac{Y_0\pm\frac{L}{2}}{r})$ (solid), for $\frac{L}{r}=10$, and $Y_0=0$. }
\label{fig:gamma function}
\end{figure}
\end{center}

The plots of  $\gamma$ and $\Gamma$ in Fig.~\ref{fig:gamma function}, shows a modest deviation between them for the case of perpendicular rods.  

\subsubsection{Rods at a Finite Angle \label{sec:Far-field finite angle}}

Figure~\ref{fig:Uss approx and exact} shows that the $X$- and $Y$-dependence is captured perfectly by Eq.~\ref{eq:Uss,app}, for perpendicular rods.  We next ask if we can capture the angle-dependence as well. It's tempting to use Eq.~\ref{eq:Uss,app} with $G(\frac{X \left|\sin \theta\right|}{r})$ and $G(\frac{Y \left|\sin \theta\right|}{r})$. We found that this approximation, while not perfect, is indeed acceptable since significant deviations are limited to 
when the rods are simultaneously not perpendicular and $X$ and $Y$ are near zero, i.e., when the ends of the rods are near each other. This is a very specific situation, and away from this the fit is almost perfect, as shown in Fig.~\ref{fig:Uss fit, 90deg}. The figure also shows Eq.~\ref{eq:Uss,app} with function $G$ replaced by its quasi-linear version, $g$. 




\begin{center}
\begin{figure}[H]
\includegraphics[width=1\linewidth]{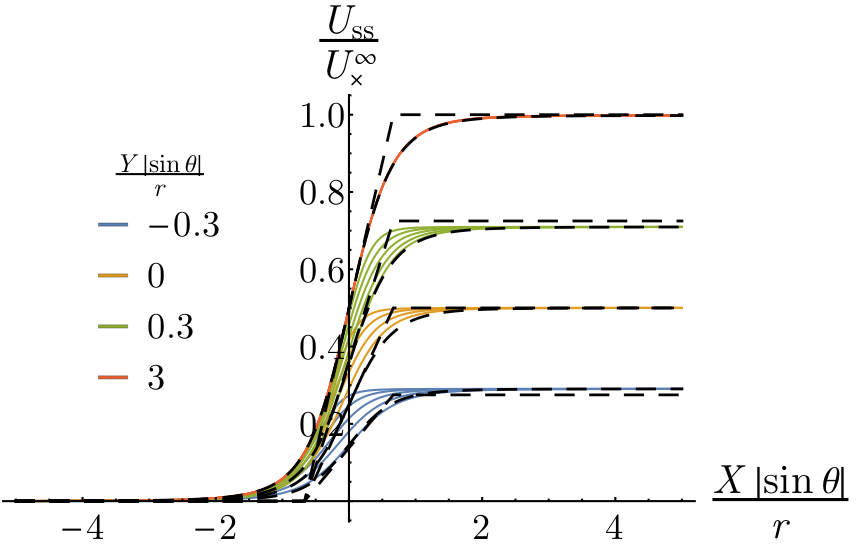} 
\caption{Approximations from Eq.~\ref{eq:Uss,app}, modified for a finite angle case with $\left|\sin \theta\right|$ for $X$ and $Y$.  Both original  ($G$) , and quasi-linear ($g$) corrections are shown as dashed lines. Exact results are given solid lines. $\frac{L}{r}=25$.}
\label{fig:Uss fit, 90deg}
\end{figure}
\end{center}

Hence, the potential for finite length, non-parallel rods in the far-field regime can be written as 

\begin{equation}\label{eq:FF U w/o angles adjusted}
U = U_{\times}^{\infty} \gamma\left(\frac{(X_0 \pm \frac{L}{2}) \left|\sin \theta\right|}{r}, \frac{(Y_0 \pm \frac{L}{2}) \left|\sin \theta\right|}{r} \right).
\end{equation}



\subsubsection{Parallel Rods}

The parallel rod-limit is substantially different from the one discussed above. The interaction potential can in fact be approximately  obtained from Eq. \ref{eq:Urods parallel, limits}, by replacing  $L$ with the overlap length, $\Delta=L-|X_0-Y_0|$ (we assume $\hat{n}_x=\hat n_y$). More precisely, it can be calculated by integration for two semi-infinite rods: 

\begin{equation*}
U(\Delta) = \frac{-A\sigma_1 \sigma_2}{\pi^2} \int_{-\infty}^{\Delta} \int_0^\infty \frac{dX \, dY}{\left[(Y-X)^2 + r^2\right]^3}.
\end{equation*}

From this integral, one can obtain the dependence of the interaction energy on the relative displacement of two finite rods, and it is indeed nearly proportional to the overlap, as shown in Fig.~\ref{fig:parallel, FF, exact and approx}. We can now modify our finite size correction $\gamma$ in such a way that in the limit of parallel rods it is also proportional to the  overlap:


\begin{center}
\begin{figure}[H]
\includegraphics[width=1\linewidth]{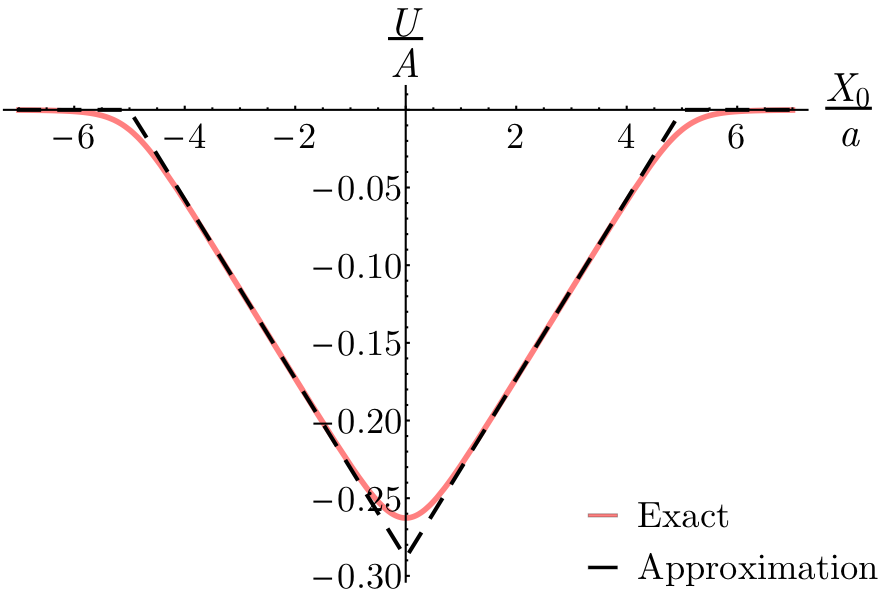} 
\caption{Comparison of the exact far-field solution for parallel rods  overlapping by $\frac{L-|X_0|}{a}$ with the approximation derived from Eq.~\ref{eq:Urods parallel, limits}, for $\frac{L}{a}=5$ and $\frac{r}{a}=1.05$.}
\label{fig:parallel, FF, exact and approx}
\end{figure}
\end{center}

\begin{equation}\label{FF U with angles adjusted}
U = \frac{-V_0 \, \gamma\left(X_\pm, Y_\pm \right)}{(\left| \sin\theta \right| + 2.35 \frac{r}{L}) \, r^4},
\end{equation}

where 
$$X_\pm = \left(X_0 \pm \frac{L}{2}\right)\left(\frac{\left|\sin \theta \right|}{r+a}+\frac{4 \left| \cos \theta \right|}{3L}\right) - \frac{4Y_0 \cos \theta}{3L} $$ 
$$Y_\pm = \left(Y_0 \pm \frac{L}{2}\right)\left(\frac{\left|\sin \theta \right|}{r+a}+\frac{4 \left| \cos \theta \right|}{3L}\right) - \frac{4X_0 \cos \theta }{3L}.$$

\begin{center}
\begin{figure}[htbp]
\includegraphics[width=1\linewidth]{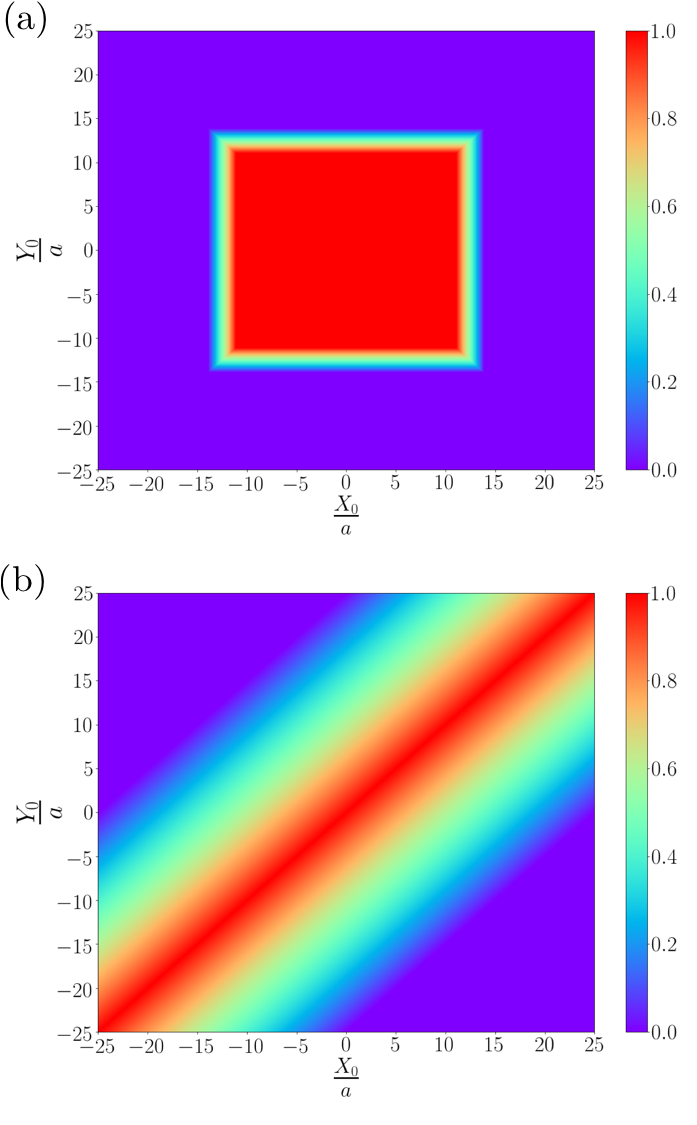} 
\caption{Heat map of $\gamma(X_\pm, Y_\pm)$ with $\frac{L}{a}=25$, for (a) $\theta=\frac{\pi}{2}$ and (b) $\theta=0$.}
\label{fig:gamma for 0 and 90 deg}
\end{figure}
\end{center}

A 2D heat map of $\gamma(X_\pm, Y_\pm)$ is shown in Fig.~\ref{fig:gamma for 0 and 90 deg} for $\theta=0^\circ$ and $\theta=90^\circ$.

\subsection{Final Form of the Attractive Energy \label{sec:Final Form of Potential}}

The final step is to combine the solution for finite thickness and no end effects (Eq.~\ref{eq:Urods, final}), with the far-field finite-length result, $\gamma(X_\pm, Y_\pm)$. 
One possibility is to simply use $\gamma$ as a corrective factor to Eq. \ref{eq:Urods, final}: $U_{\mathrm{rods}}\gamma$. This satisfies the case of finite rods in the far-field or away from the ends of the rods, but there is a "shadow effect" in the near-field that makes the rods appear longer than they are. This shadow effect comes from the divergence of the $(\frac{1}{r-a})$ factors in $U_{\mathrm{rods}}$, and it is not physical when either $|X_0|$ or $|Y_0|$ is greater than $\frac{L}{2}$ for perpendicular rods, or $L$ for parallel rods. In addition, $a$ is, in general, a function of $X$ and $Y$. To account for non-uniform shaped rods and fix the shadow effect, we have included a function $\gamma_a$ as a prefactor to $a$ in the linear term of the denominator. This makes it possible to define the shape of the rods' terminal  by causing the diameter to shrink to zero in a specific way, effectively forming a ``cap". To account for the difference in the maximum $|X_0|$ or $|Y_0|$ that leads to a collision for different angles, $\gamma_a$ was designed to equal one, and to begin to decay to zero as $\gamma$ goes to zero. Using this $\gamma_a$ will ensure that the rod diameter is uniform for most of its length, but will be zero when the rods are in positions where they should not collide as $r$ decreases. For simplicity we have used $\gamma_a = \gamma(X_\pm^a, Y_\pm^a)$, with $X_\pm^a$ and $Y_\pm^a$ defined below, but it can be replaced by other functions to describe rod-like objects of different shapes (e.g. ellipsoids). With the $\gamma_a$ correction, the potential has a physical  global behavior.
The final form of the attractive potential energy, $U_{\mathrm{vdW}}$, is shown below. Figure~\ref{fig:Uvdw heatmap} shows a two-dimensional plot of the magnitude of $U_{\mathrm{vdW}}$.  


\begin{widetext}
\begin{equation}\label{eq:Uattr final}
\boxed{U_{\mathrm{vdW}}(X_0, Y_0, r, L,\theta) = \frac{-A \pi a^4 \, \gamma\left(X_\pm, Y_\pm \right)}{32 \, (\left| \sin \theta \right|+2.35 \frac{\sqrt{r(r-\gamma_a a)}}{L}) \, (r- \gamma_{a} a) \, (r+0.12 \, a)^3}}
\end{equation}

\begin{align*}
\gamma(x\pm, y\pm)&=\mathrm{min}\left\{\left[g\left(x_+ \right)-g\left(x_-\right) \right], \left[g\left(y_+ \right)-g\left(y_-\right) \right]\right\}&g(x)&=\frac{1}{2} \, \mathrm{sgn}\left(x\right) \, \mathrm{min}\left\{ 1, \frac{3}{2}\left|x\right| \right\}&\gamma_a&=\gamma\left(X_\pm^a, Y_\pm^a \right)
\end{align*}
{\small
\begin{align*}
X_\pm&=\left(X_0 \pm \frac{L}{2}\right)\left(\frac{\left|\sin \theta \right|}{r+a}+\frac{4 \left| \cos \theta \right|}{3L}\right) - \frac{4Y_0 \cos \theta}{3L}&Y_\pm&=\left(Y_0 \pm \frac{L}{2}\right)\left(\frac{\left|\sin \theta \right|}{r+a}+\frac{4 \left| \cos \theta \right|}{3L}\right) - \frac{4X_0 \cos \theta}{3L}
\end{align*}
}
\begin{align*}
X_\pm^a &= \frac{(X_0\pm \frac{L}{2})}{a} \pm \frac{2(r+a)(L\mp 2Y_0\cos \theta)}{4a(r+a) \left|\cos \theta\right| + 3La \left|\sin \theta\right|} \mp 1.4&Y_\pm^a &= \frac{(Y_0\pm \frac{L}{2})}{a} \pm \frac{2(r+a)(L\mp 2X_0\cos \theta)}{4a(r+a) \left|\cos \theta\right| + 3La \left|\sin \theta\right|} \mp 1.4
\end{align*}
\end{widetext}





\subsection{Repulsion, Force field, and Torques. }

\begin{figure}[hbt]
\includegraphics[width=1\linewidth]{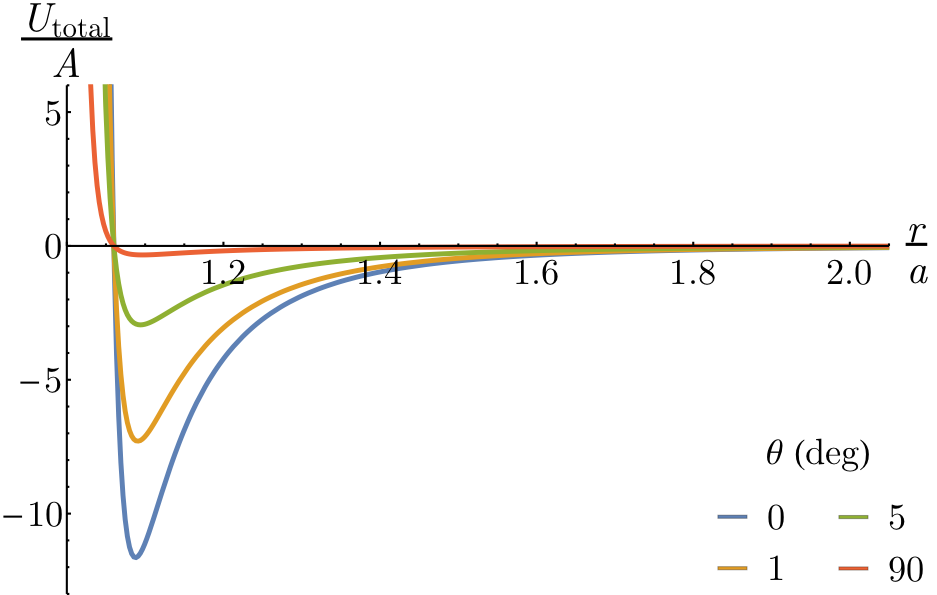} 
\caption{$U_{\mathrm{vdW}}$ with power-law repulsion from Eq.~\ref{eq:Full Potential}, plotted for several angles with $\frac{X_0}{a}=\frac{Y_0}{a}=0$, $\frac{L}{a}=25$, and $\xi=0.12$.}
\label{fig:Utotal with pow}
\end{figure}

For practical use of the obtained results, one needs to combine the vdW attraction with certain model repulsion. The latter may be system specific, as in the case of nanorods and colloids stabilized electrostatically, or with  ligands. Nevertheless, as long as the repulsion has hard core character, the details are not very  important. Here we propose two versions of a full potential. They combine our vdW result $U_{\mathrm{vdW}}$ with either algebraic or exponential repulsion: 

\begin{equation} \label{eq:Full Potential}
U_\mathrm{total} = U_{\mathrm{vdW}} \left(1-\left(\frac{\xi}{r-\gamma_a a} \right)^2\right)\\
\end{equation}

\begin{equation}
U_{\mathrm{total}} = U_{\mathrm{vdW}} \left( 1-e^{(1-\frac{r-\gamma_a a}{\xi})} \right) 
\end{equation}
Figure~\ref{fig:Utotal with pow} shows the  algebraic  version of this combined potential, Eq. \ref{eq:Full Potential}.

\begin{figure}[hbt]
\includegraphics[width=1\linewidth]{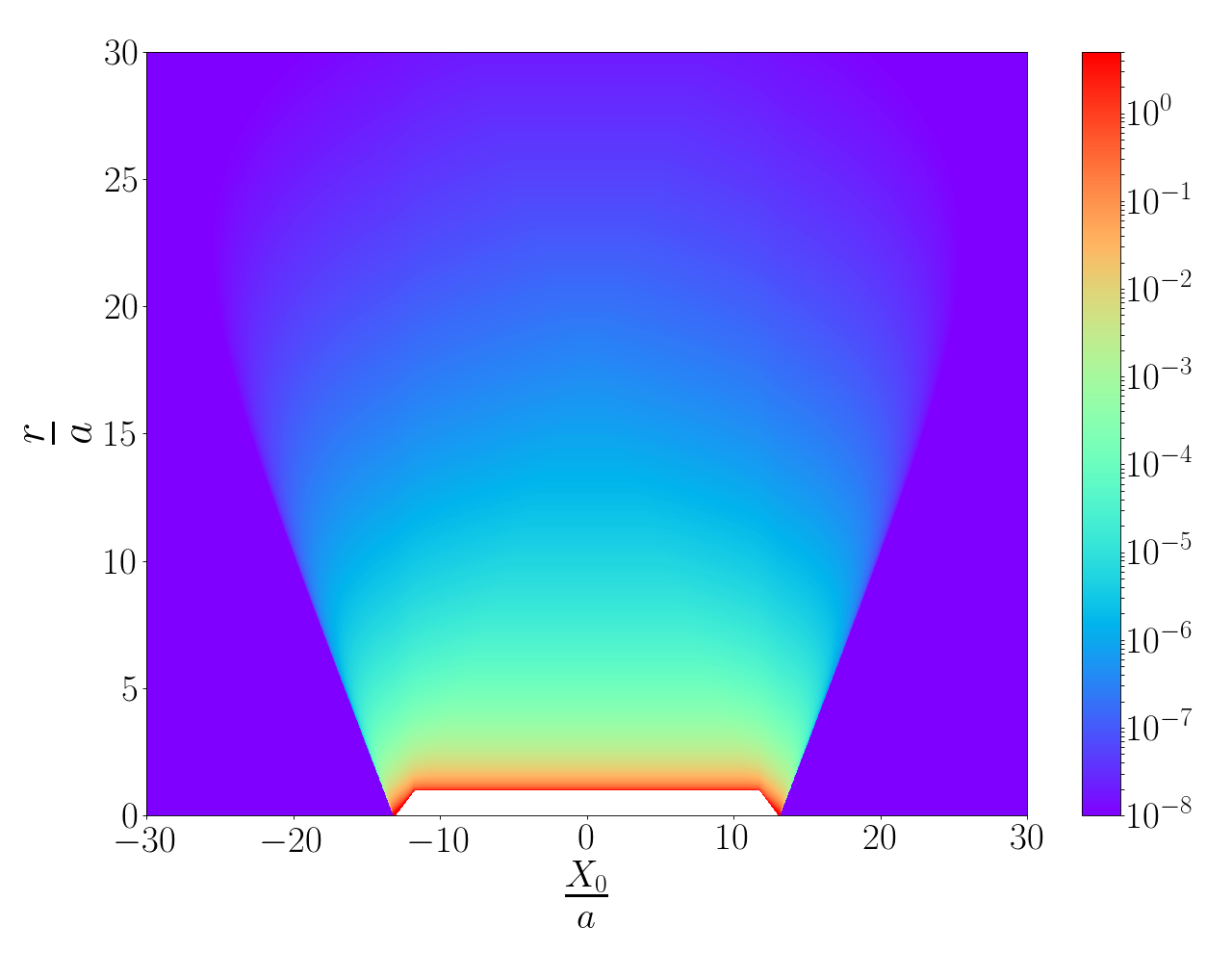} 
\caption{Two-dimensional plot of $\frac{\left|U_{\mathrm{vdW}}\right|}{A}$ with $\frac{Y_0}{a}=0$, $\theta=\frac{\pi}{2}$, and $\frac{L}{a}=25$.}
\label{fig:Uvdw heatmap}
\end{figure}

Another important practical aspect of application of our results to real simulations  (in particular, Molecular Dynamics) is the need to derive the force fields and torques from the potential. This is done in Appendix~\ref{sec:Forces and Torques}. The final results are presented  below. 

Here centers of rods $1$ and $2$ are $\vec{X}_c$ and $\vec{Y}_c$, respectively, with  indexes $1$ and $2$  selected by condition $|X_0|\leq|Y_0|$. The force acting on rod $1$, $\vec{F}_1$, is


\begin{align}\label{eq:Force}
\vec{F}_1 = U_{\mathrm{vdW}} \Biggl[ \frac{\cos\theta}{\sin^2\theta}& \Biggl( \frac{3\lambda}{4}  \vec{f}' + \frac{1}{\gamma L}\vec{f}'' \Biggr) \\ \nonumber &+ \left( \lambda + \frac{3}{r+ 0.12 a} \right) \hat{n}_z \Biggr],
\end{align}

where

\begin{scriptsize}
\begin{align*}
&\begin{aligned}
\vec{f}' &= \Biggl[ \Theta\left( \frac{2}{3}-|Y_+^a| \right) - \Theta \left( \frac{2}{3}-|Y_-^a| \right) \Biggr]\Biggl[ \left( \frac{1}{\left|\cos\theta\right| + \frac{3L}{4(r+a)} \left|\sin\theta\right|} + 1 \right) \hat{n}_x \\ 
&\quad + \left(\frac{1}{\cos\theta} + \frac{ \cos\theta}{\left|\cos\theta\right| + \frac{3L}{4(r+a)} \left|\sin\theta\right|} \right) \hat{n}_y \Biggr]
\end{aligned} \\ \\ 
&\begin{aligned}
\vec{f}'' &= \Biggl[ \Theta\left(\frac{2}{3}-|Y_+|\right) - \Theta\left(\frac{2}{3}-|Y_-|\right) \Biggr]\Biggl[ \left(1 + \left|\cos\theta\right| + \frac{3L \left|\sin\theta\right|}{4(r+a)} \right) \hat{n}_x \\ &\quad + \left(\cos\theta + \frac{\left|\cos\theta\right|}{\cos\theta} + \frac{3L \left|\sin\theta\right|}{4\cos\theta(r+a)} \right) \hat{n}_y \Biggr]
\end{aligned} \\ \\ 
&\begin{aligned}
\lambda = \frac{1}{r- \gamma_a a} \left( 1 + \frac{1}{\frac{L \, \left|\sin \theta \right|}{a}+2} \right),
\end{aligned}
\end{align*}
\end{scriptsize}

and $\Theta(x)$ is the Heaviside step function.

The force acting on  rod $2$ is, of course, $-\vec{F}_1$. The torques on the rods are found to be

\begin{equation}\label{eq:Torque}
\begin{aligned}
\vec{\boldsymbol{\tau}}_{1} &= -X_0 \, \hat{n}_x \times \vec{F}_1 - \frac{U_{\mathrm{vdW}} \, \cot\theta}{1 + \frac{2.35}{L \left|\sin\theta\right|}\sqrt{r \left(r-\gamma_a a\right)}} \: \hat{n}_z \\ 
\vec{\boldsymbol{\tau}}_{2} &= Y_0 \, \hat{n}_y \times \vec{F}_1 + \frac{U_{\mathrm{vdW}} \, \cot\theta}{1 + \frac{2.35}{L \left|\sin\theta\right|}\sqrt{r \left(r-\gamma_a a\right)}} \: \hat{n}_z.
\end{aligned}
\end{equation}

\section{Conclusion}

In this work, we have constructed a compact analytic description of van der Waals interaction between two identical rods. Our model is applicable to metallic nanorods, rod-like colloids, and multi-wall carbon nanotubes.  The resulting potential, given by Eq.~\ref{eq:Full Potential}, can be used directly, e.g. for Monte Carlo simulations. The force fields and torques derived from it are practical for Molecular or Brownian Dynamics. Note that for the problems in which the end effects are not essential, a simpler version of the  potential Eq. \ref{eq:Urods, final} can be used. In that case, our interpolative formula is indistinguishable from the exact result, for any distance and angle between the rods. For instance, this potential can be used to include van der Waals effects in Onsager theory of Isotropic-Nematic liquid crystal transition \cite{onsager1949effects}.

Limitations of our model include the usual neglect for retardation and many-body effects, such as  screening. On the other hand, our approach is easily generalizable for the cases beyond simple cylindrical geometry. Namely, the shape  of interacting objects is primarily determined by the function $\gamma_a$ in  Eq.~\ref{eq:Full Potential}. Thus, by changing this single function one can describe other axially symmetric elongated shapes, e.g.  ellipsoids.  Finally, one can combine our approach with the theory of Zhbanov {\it et al.} \cite{zhbanov2010van,pogorelov2011universal}, making it suitable both for single-wall and multi-wall  carbon nanotubes of arbitrary thickness.


\begin{acknowledgments}
This research was carried out at  Center for Functional Nanomaterials, which is a U.S. DOE Office of Science Facility, at Brookhaven National Laboratory under Contract No. DE-SC0012704.
\end{acknowledgments}

\appendix

\section{Derivation of the Asymptotes of $U_{\times}$ \label{sec:Appendix Asymptote Derivation}}

The interaction potential energy for two rods is found using

\begin{equation*}\label{eq:Vol Interaction Appendix}
U = \frac{-A}{\pi^2} \int \frac{d^3 \vec{r}_1 \, d^3 \vec{r}_2}{|\vec{r}_1-\vec{r}_2|^6}
\end{equation*}

For thin rods the potential becomes


\begin{equation}
U=\frac{-A \, \sigma_1 \sigma_2}{\pi^2 \, r^4 \left| \sin \theta \right|} \int_{x_1}^{x_2} \int_{y_1}^{y_2} \frac{dy \, dx}{(1+x^2+y^2)^3} 
\end{equation}

where $x_{1/2}$ and $y_{1/2}$ are the endpoints of the rods, measured along the $\hat{n}_x$ and $\hat{n}_y$ axes. The interaction between skew rods with infinite length in the far-field ($U_\times^\infty$) and for an infinite length rod perpendicular to a semi-infinite rod ($U_{\mathrm{s}}$) are found as

\vspace{5mm}

\begin{equation}\label{eq:Uinf Appendix}
U_\times^\infty=\frac{-A \, \sigma_1 \sigma_2}{\pi^2 \, r^4 \left| \sin \theta \right|} \int_{-\infty} ^\infty \int_{-\infty} ^\infty \frac{dy \, dx}{(1+x^2+y^2)^3} =  \frac{-A \, \sigma_1 \sigma_2}{2 \pi r^4 \left| \sin\theta \right|}
\end{equation}

\vspace{5mm}

\begin{eqnarray}\label{eq:Uis}
U_{\mathrm{s}}=\frac{-A \, \sigma_1 \sigma_2}{\pi^2 \, r^4 \left| \sin \theta \right|} \int_{-\infty} ^{\frac{X}{r}} \int_{-\infty} ^\infty \frac{dy \, dx}{(1+x^2+y^2)^3} \nonumber \\ = U_\times^\infty \left[ \frac{3\frac{X}{r} + 2(\frac{X}{r})^3}{4(1+(\frac{X}{r})^2)^{3/2}}+\frac{1}{2} \right]  
\end{eqnarray}

\vspace{5mm}






$U_\times^\infty$ can be used to find the interaction between two infinite, finite diameter rods as

\begin{equation}
U_{\times} = \int \int \frac{-A \, d\sigma_1 d\sigma_2}{2 \pi \left| \sin \theta \right| (\vec{z}_1 - \vec{z}_2)^4} 
\end{equation}

\begin{center}
\begin{figure}[h]
\includegraphics[scale=0.5]{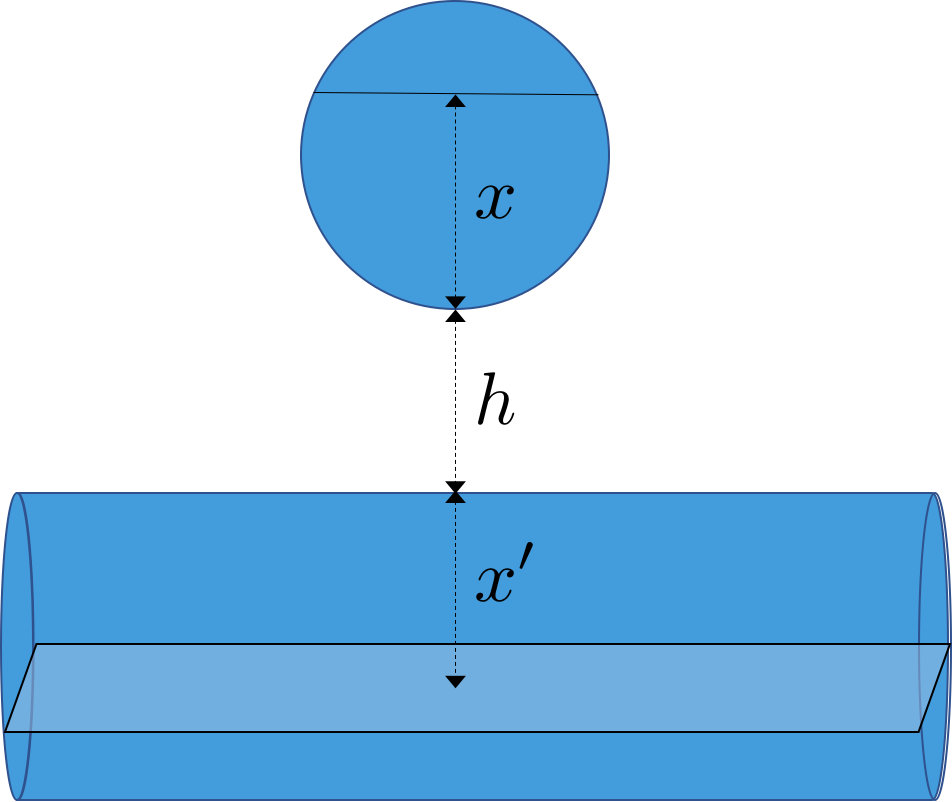} 
\caption{Distance between cross-sections of the rods is given by $h+x+x'$, and is completely in the $\hat{n}_z$ direction.}
\label{fig:finite diameter model}
\end{figure}
\end{center}

\begin{center}
\begin{figure}[h]
\includegraphics[scale=0.5]{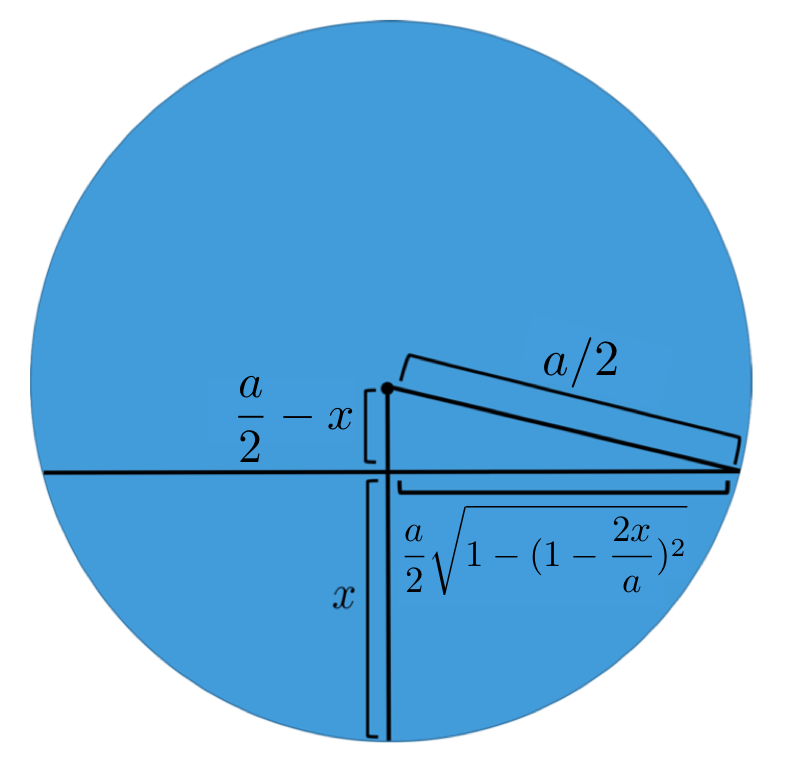} 
\caption{Cross section of one rod showing the width $w(x)$.}
\label{fig:finite diameter variables}
\end{figure}
\end{center}

The distance between infinitesimal cross-sections depends only on the $\hat{n}_z$ direction, and is given by $h+x+x'$, as shown in Fig.~\ref{fig:finite diameter model}, where $h=r-a$ is the surface-to-surface distance between the rods.

\begin{equation*}
U_{\times} = \int_0^a \int_0^a \int_0^{w(x')} \int_0^{w(x)} \frac{-A}{2 \pi \left| \sin \theta \right|} \frac{dy \, dy' \, dx \, dx'}{(h+x+x')^4} 
\end{equation*}


where $x$ and $x'$ are defined as the distance from the edge nearest the opposite rod to a cross-section along the length of the rod, as shown in Figs.~\ref{fig:finite diameter model} and \ref{fig:finite diameter variables}. $w(x)=~a\sqrt{1-(1-\frac{2x}{a})^2}$ is the width of the rod at a cross-section $x$ from the edge.

After integration over $y$ and $y'$, the energy for a pair of infinite rods with finite diameter is


\begin{equation*}
U_{\times} = \frac{-2A a^2}{\pi \left|\sin\theta\right|} \int_0 ^{a} \int_0 ^{a}  \frac{ \sqrt{\frac{x}{a}(1-\frac{x\vphantom{'}}{a})}\: \sqrt{\frac{x'}{a}(1-\frac{x'}{a})}}{(h+x+x')^4} dx dx'
\end{equation*}

With the substitution $u=\frac{x}{a}$ it becomes

\begin{equation*}
U_{\times} = \frac{-2A}{\pi a \left|\sin\theta\right|} \int_0 ^{a} \left( \int_0 ^{1}  \frac{ \sqrt{u \, (1-u)}}{(\frac{H}{a}+u)^4} du \right) \sqrt{\frac{x'}{a}(1-\frac{x'}{a})} \: dx'
\end{equation*}

\begin{center}
\begin{figure}[h]
\includegraphics[scale=0.35]{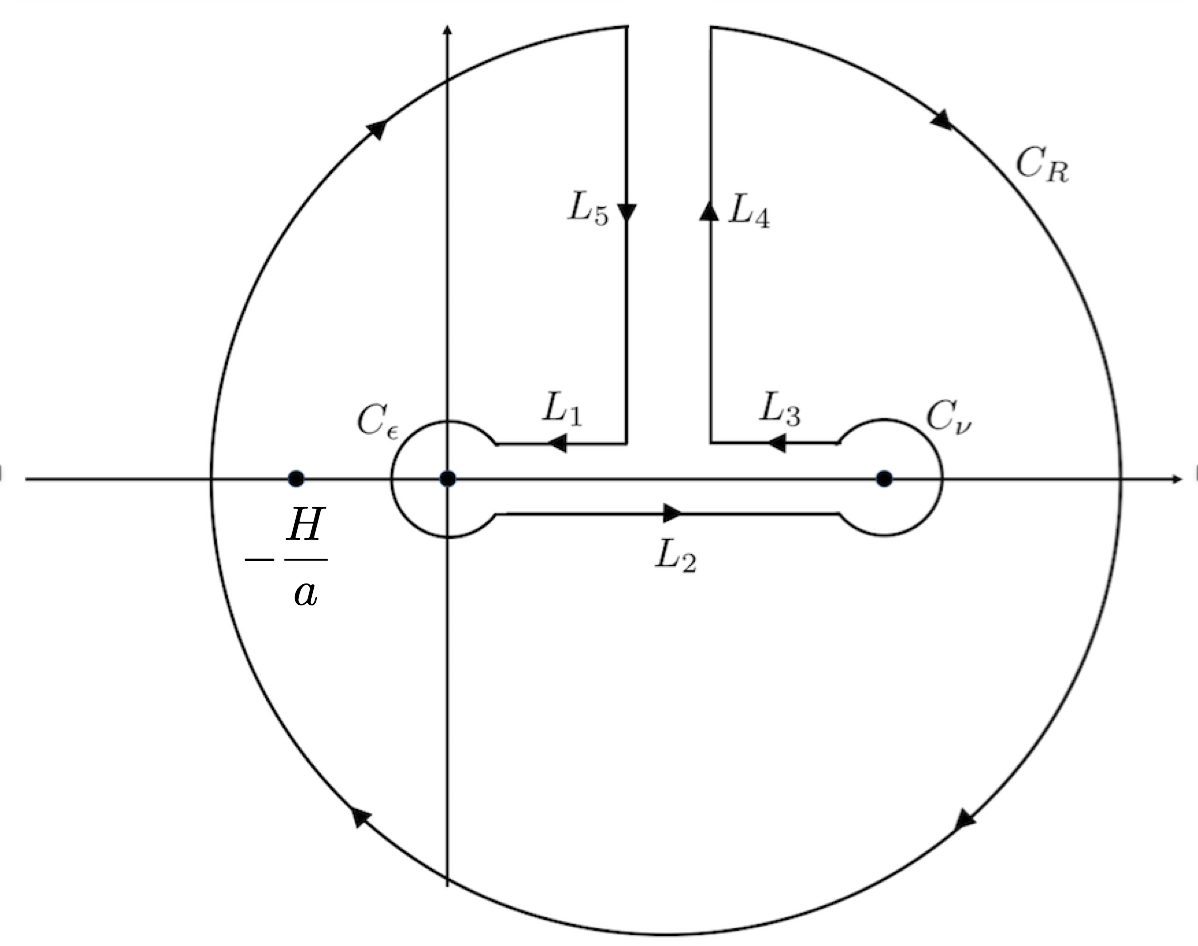} 
\caption{Contour of first integral with fourth-order pole at $u=-\frac{H}{a}$.}
\label{fig:Contour Labels}
\end{figure}
\end{center}

The first integral is done using contour integration with the contour shown in Fig.~\ref{fig:Contour Labels}. There is a pole at $u=-\frac{H}{a}$.

The result of the contour integral leads to 

\begin{equation*}
U_{\times} = \frac{-A}{8a \left|\sin\theta\right|} \int_0 ^{a} \frac{\sqrt{\frac{x'}{a}(1-\frac{x'}{a})} \, (2\frac{x'}{a} + 2\frac{h}{a} + 1)}{[(\frac{x'}{a}+\frac{h}{a})(\frac{x'}{a}+\frac{h}{a}+1)]^{\frac{5}{2}}} \: dx'
\end{equation*}

To evaluate this integral, we chose to break it into two approximations, $h \gg a$ and $h \ll a$. 

In the far-field, $h \gg a$, we take the lowest-order approximation, and arrive at 

\begin{equation*}
U_{\times}^\infty \approx \frac{-A a^3}{4 h^4 \left| \sin \theta \right|} \int_0^a \sqrt{\frac{x'}{a} \left(1-\frac{x'}{a} \right)} \, dx' = \frac{-A \pi a^4}{32 \left| \sin \theta \right| r^4}
\end{equation*}

In the last step, $h$ is replaced with $r$.

In the near-field, $h \ll a$, taking care with the divergence that occurs at the lower limit of the integral, we expand the integrand for both $h \ll a$ and $x' \ll a$.

Taking both approximations to lowest-order and rewriting the integrand in terms of $\frac{x'}{h}$ leads to

\begin{equation*}
U_{\times}^0 \approx \frac{-Aa}{8h^2 \left| \sin \theta \right|} \int_0^\infty \frac{\sqrt{\frac{x'}{h}}}{(1+\frac{x'}{h})^{5/2}} \, dx' = \frac{-A a}{12 \, \left| \sin \theta \right| (r-a)}
\end{equation*}

\section{Forces and Torques \label{sec:Forces and Torques}}

\subsection{Force Between Rods \label{sec:Forces}}

In this section we would like to evaluate the forces and torques that the rods exert on one another. To calculate the forces between the rods, we need to take the gradient of the potential energy in the $\hat{n}_x$, $\hat{n}_y$, and $\hat{n}_z$ directions. The $X$-$Y$ system is, in general, non-orthogonal, and the force can be derived by recognizing that, for some $\alpha$, $\beta$, and $\delta$, 

\begin{align*}
\vec{\boldsymbol{\nabla}}U &= \alpha \hat{n}_x + \beta
\hat{n}_y + \delta \hat{n}_z
\end{align*}
\begin{align*}
\hat{n}_x \cdot \hat{n}_x&=1&\hat{n}_y \cdot \hat{n}_y&=1&\hat{n}_z \cdot \hat{n}_z&=1
\end{align*}
\begin{align*}
\hat{n}_x \cdot \hat{n}_z&=0&\hat{n}_y \cdot \hat{n}_z&=0&\hat{n}_x \cdot \hat{n}_y&=\cos\theta
\end{align*}

Using these equations leads to a system of equations in $\alpha$, $\beta$, and $\delta$.

\begin{align*}
\vec{\boldsymbol{\nabla}}U \cdot \hat{n}_x &= \frac{\partial U}{\partial X_0} = \alpha + \beta \cos \theta \\ 
\vec{\boldsymbol{\nabla}}U \cdot \hat{n}_y &= \frac{\partial U}{\partial Y_0} = \alpha \cos \theta + \beta \\ 
\vec{\boldsymbol{\nabla}}U \cdot \hat{n}_z &= \frac{\partial U}{\partial r} = \delta
\end{align*}

The system can be solved to give the gradient in $XYZ$ coordinates. We will consider that $|X_0|\leq|Y_0|$ is always true, and call the rod with center of mass position $\vec{X}_c$, rod 1, and the rod with center of mass position $\vec{Y}_c$, rod 2. The $Y$-component of the force is made negative so that this always gives the force on rod 1. 

\begin{small}
\begin{widetext}
\begin{equation}\label{eq:Force in XYZ system}
\vec{F}=-\vec{\boldsymbol{\nabla}} U_{\mathrm{vdW}} = -\left[ \frac{1}{\sin^2 \theta} \left( \frac{\partial U_{\mathrm{vdW}}}{\partial X_0}- \cos \theta \, \frac{\partial U_{\mathrm{vdW}}}{\partial Y_0}  \right) \, \hat{n}_x - \frac{1}{\sin^2 \theta} \left( \frac{\partial U_{\mathrm{vdW}}}{\partial Y_0}- \cos \theta \, \frac{\partial U_{\mathrm{vdW}}}{\partial X_0}  \right) \, \hat{n}_y + \left( \frac{\partial U_{\mathrm{vdW}}}{\partial r} \right) \hat{n}_z  \right]
\end{equation}
\end{widetext}
\end{small}

Keeping only the dominant terms, and including a factor to tune the fit for different angles, the derivatives are approximated as below.

\begin{align*}
\frac{\partial U_{\mathrm{vdW}}}{\partial X_0} &\approx U_{\mathrm{vdW}} \left[a \lambda \frac{\partial \gamma_a}{\partial X_0} + \frac{1}{\gamma}\frac{\partial \gamma}{\partial X_0} \right] \\ 
\frac{\partial U_{\mathrm{vdW}}}{\partial Y_0} &\approx U_{\mathrm{vdW}} \left[a \lambda \frac{\partial \gamma_a}{\partial Y_0} + \frac{1}{\gamma}\frac{\partial \gamma}{\partial Y_0} \right] \\ 
\frac{\partial U_{\mathrm{vdW}}}{\partial r} &\approx -U_{\mathrm{vdW}} \left[ \lambda + \frac{3}{r+\epsilon a} \right] \\ 
\lambda &= \frac{1}{r- \gamma_a a} \left( 1 + \frac{1}{\frac{L \, \left|\sin \theta \right|}{a}+2} \right)
\end{align*} \\

With the derivatives of $\gamma$ and $\gamma_a$ given by

\begin{footnotesize}
\begin{align*}
\frac{\partial \gamma}{\partial X_0} &= \frac{-\cos \theta}{L} \left[ \Theta\left(\frac{2}{3}-|Y_+|\right) -  \Theta\left(\frac{2}{3}-|Y_-|\right) \right] \\ \\ 
\frac{\partial \gamma}{\partial Y_0} &= \left(\frac{3 \left|\sin \theta \right|}{4 (r+a)}+\frac{\left|\cos \theta \right|}{L} \right) \left[ \Theta\left(\frac{2}{3} - |Y_+|\right) -  \Theta\left(\frac{2}{3} - |Y_-|\right) \right] \\ \\ 
\frac{\partial \gamma_a}{\partial X_0} &= \left(\frac{-3 \cos \theta}{4a \left|\cos \theta \right| + \frac{3La}{r+a} \left|\sin \theta \right|}\right) \left[\Theta\left( \frac{2}{3}-|Y_+^a| \right) - \Theta \left( \frac{2}{3}-|Y_-^a| \right) \right] \\ \\ 
\frac{\partial \gamma_a}{\partial Y_0} &= \frac{3}{4a} \left[\Theta\left(\frac{2}{3}-|Y_+^a| \right) - \Theta \left( \frac{2}{3}-|Y_-^a| \right) \right]
\end{align*}
\end{footnotesize}

\vspace{5mm}

Altogether, the force is

\begin{align}\label{eq:Appendix Force}
\vec{F}_1 = U_{\mathrm{vdW}} \Biggl[ \frac{\cos\theta}{\sin^2\theta}& \Biggl( \frac{3\lambda}{4}  \vec{f}' + \frac{1}{\gamma L}\vec{f}'' \Biggr) \\ \nonumber &+ \left( \lambda + \frac{3}{r+ 0.12 a} \right) \hat{n}_z \Biggr],
\end{align} \\

where

\begin{scriptsize}
\begin{align*}
&\begin{aligned}
\vec{f}' &= \Biggl[ \Theta\left( \frac{2}{3}-|Y_+^a| \right) - \Theta \left( \frac{2}{3}-|Y_-^a| \right) \Biggr]\Biggl[ \left( \frac{1}{\left|\cos\theta\right| + \frac{3L}{4(r+a)} \left|\sin\theta\right|} + 1 \right) \hat{n}_x \\ 
&\quad + \left(\frac{1}{\cos\theta} + \frac{ \cos\theta}{\left|\cos\theta\right| + \frac{3L}{4(r+a)} \left|\sin\theta\right|} \right) \hat{n}_y \Biggr]
\end{aligned} \\ \\ 
&\begin{aligned}
\vec{f}'' &= \Biggl[ \Theta\left(\frac{2}{3}-|Y_+|\right) - \Theta\left(\frac{2}{3}-|Y_-|\right) \Biggr]\Biggl[ \left(1 + \left|\cos\theta\right| + \frac{3L \left|\sin\theta\right|}{4(r+a)} \right) \hat{n}_x \\ &\quad + \left(\cos\theta + \frac{\left|\cos\theta\right|}{\cos\theta} + \frac{3L \left|\sin\theta\right|}{4\cos\theta(r+a)} \right) \hat{n}_y \Biggr],
\end{aligned}
\end{align*}
\end{scriptsize}

and $\Theta(x)$ is the Heaviside step function.\\ \\

\subsection{Torque on the Rods \label{Torques}}

Given the forces from Eq.~\ref{eq:Force in XYZ system}, we can write down the torque. The derivative of $U_{\mathrm{vdW}}$ is approximated as

\begin{equation*}
\frac{\partial U_{\mathrm{vdW}}}{\partial \theta} \approx - \frac{U_{\mathrm{vdW}} \, \cot\theta}{1 + \frac{2.35}{L \left|\sin\theta\right|}\sqrt{r \left(r-\gamma_a a\right)}}
\end{equation*}

The torque experienced by rod 1 from rod 2, $\vec{\boldsymbol{\tau}}_{1}$, comes from the force $\vec{F}_{1}=-\vec{F}_{2}=\vec{F}$. Recalling the definition

\begin{align*} 
\hat{n}_x \times \hat{n}_y&=\sin \theta \, \hat{n}_z, \\ 
\end{align*}

the torques are given by

\begin{equation}\label{eq:Appendix Torque}
\begin{aligned}
\vec{\boldsymbol{\tau}}_{1} &= -X_0 \, \hat{n}_x \times \vec{F} - \frac{U_{\mathrm{vdW}} \, \cot\theta}{1 + \frac{2.35}{L \left|\sin\theta\right|}\sqrt{r \left(r-\gamma_a a\right)}} \: \hat{n}_z \\ 
\vec{\boldsymbol{\tau}}_{2} &= Y_0 \, \hat{n}_y \times \vec{F} + \frac{U_{\mathrm{vdW}} \, \cot\theta}{1 + \frac{2.35}{L \left|\sin\theta\right|}\sqrt{r \left(r-\gamma_a a\right)}} \: \hat{n}_z.
\end{aligned}
\end{equation}

\end{document}